# Characterization of Blood Pressure and Heart Rate Oscillations of POTS Patients via Uniform Phase Empirical Mode Decomposition

Justen Geddes[1], Jesper Mehlsen[2], Mette S. Olufsen[1]

1) Department of Mathematics, North Carolina State University, Raleigh, NC
2) Surgical Pathophysiology Unit, Rigshospitalet, Copenhagen, Denmark

*Abstract— **Objective:** Postural Orthostatic Tachycardia Syndrome (POTS) is associated with the onset of tachycardia upon postural change. The current diagnosis involves the measurement of heart rate (HR) and blood pressure (BP) during head-up tilt (HUT) or active standing test. A positive diagnosis is made if HR changes with more than 30 bpm (40 bpm in patients aged 12-19 years), ignoring all of the BP and most of the HR signals. This study examines 0.1 Hz oscillations in systolic arterial blood pressure (SBP) and HR signals providing additional metrics characterizing the dynamics of the baroreflex. **Methods:** We analyze data from 28 control subjects and 28 POTS patients who underwent HUT. We extract beat-to-beat HR and SBP during a 10 min interval including 5 minutes of baseline and 5 minutes of HUT. We employ Uniform Phase Empirical Mode Decomposition (UPEMD) to extract 0.1 Hz stationary modes from both signals and use random forest machine learning and k-means clustering to analyze the outcomes. **Results** show that the amplitude of the 0.1 Hz oscillations is higher in POTS patients and that the phase response between the two signals is shorter (p < 0.005). **Conclusion:** POTS is associated with an increase in the amplitude of SBP and HR 0.1 Hz oscillation and shortening of the phase between the two signals. **Significance:** The 0.1 Hz phase response and oscillation amplitude metrics provide new markers that can improve POTS diagnostic augmenting the existing diagnosis protocol only analyzing the change in heart rate.*

## I. INTRODUCTION

Postural Orthostatic Tachycardia Syndrome (POTS) is associated with the presence of chronic (more than six months) tachycardia measured during head-up tilt (HUT) or active standing combined with a history of orthostatic intolerance [1]. POTS is a phenotype and not a specific disease, the symptoms likely can be caused by several pathophysiological mechanisms spanning from tachycardia caused by dehydration to hyperadrenergic mechanisms due to agonistic antibodies to specific adrenergic receptors [2]. Symptoms associated with POTS include dizziness, nausea, palpitations, visual blurring, and/or brain fog appearing during the transition from sitting or supine to upright position [3]. These symptoms may be mild, but they can lead to severe incapacitation [4]. Positive diagnosis for an adult (20 years or older) is defined as an increase in heart rate of more than 30 bpm within 10 minutes after onset of the HUT, whereas for children and young adults (aged 12 to 19 years) positive diagnoses is associated with a heart rate increase of more than 40 bpm [5]. Postural tachycardia is the current approach to identify POTS but as pointed by Raj and Robertson there is a need for more detailed diagnostic approaches to differentiate POTS patients with respect to pathophysiological mechanisms [6]. A first step is to determine if blood pressure and heart rate signals contain other characteristics that further describe the patients. The purpose of our study is to use signal processing to explore this approach.

An exact definition of the interval over which heart rate should be monitored does not exist. The American College of Cardiologists recommends diagnosing patients with POTS if tachycardia is observed within the first 10 minutes of postural change. This criteria was used by Wang *et al*. [13], who found that POTS patients exhibit tachycardia 5-10 min following postural change, while Kirbis *et al*. [7] argue that it is adequate to measure heart rate for 3 minutes following the postural change. These differences likely occur due to the simple one-value measure used in diagnostic criteria, highlighting the need for a more detailed protocol to analyze heart rate and blood pressure signals.

Approximately 75% of patients experiencing POTS are young women aged 20 to 40 years old [8], and the disease onset is typically induced by acute stressors, including viral illness [9], pregnancy [10], and injury [11]. For some patients, the disease onset has been observed after the administration of the Human Papillomavirus vaccine; however, a causal relationship has not been established [12]. The current diagnosis only targets the increase in heart rate, yet visual inspection of both the heart rate and blood pressure signals suggest that POTS patients experience increased oscillatory behavior at the 0.1Hz frequency associated with modulation of the baroreceptor reflex [13]. This study provides a more detailed analysis of these signals, which potentially can lead to better classification and understanding of the disease.

In healthy controls, most physiological systems operate via negative feedback keeping the system at homeostasis. A wide range of normal physiological processes oscillates at specific mean frequencies. For example, for females, the slowest frequency is the menstrual (infradian) cycle ~28 days [14], followed by circadian (~24 h) [15] and ultradian (< 24 h) cycles. Other prominent frequency responses include the baroreflex response (~0.1 Hz), respiration (~0.25 Hz), and heart rate (~1 Hz) [16, 17].

This study examines heart rate and blood pressure oscillations caused by the baroreflex feedback, which operates at a mean frequency of approximately 0.1 Hz [13]. The baroreflex is a negative feedback system increasing heart rate, vascular resistance, and cardiac contractility in response to a decrease in blood pressure sensed by the baroreceptors located in the aortic arch and carotid sinuses. In healthy adults this reflex operates along the parasympathetic and sympathetic pathways keeping heart rate and blood pressure at homeostasis [18]. Upon HUT, an immediate increase in heart rate is the first response to gravity causing an increase in flow to the lower extremities. Within 5-10 sec the resistance vessels contract due to stimulation of the adrenergic alpha-1-receptors [19]. The presence of agonistic antibodies directed at one or more parts of the baroreflex arc or changes in the elimination of transmitters in the autonomic nervous system would likely cause oscillations by enhancing the negative baroreflex feed-back loop at its resonance frequency of 0.1 Hz, while low cardiac filling (dehydration) would produce static changes.

In the data analyzed here, we observed that in addition to tachycardia in the upright position there are significant blood pressure and heart rate oscillations at ~0.1 Hz. We hypothesize that these oscillations are more prominent (with higher amplitude) in POTS patients compared to control subjects and that the phase between the HR and BP 0.1 Hz oscillations is shorter for POTS patients. To show this, we extract beat-to-beat heart rate and systolic blood pressure values over 6-10 min from 28 control and 28 POTS patients undergoing a HUT test. The signals include at least 3 minutes of data before and during HUT. To test our hypothesis, we use Uniform Phase Empirical Mode Decomposition (UPEMD) to analyze the signals [20].

Most studies analyzing heart rate and blood pressure data from POTS patients focus on characterizing the discrete change in heart rate measured in the transition from supine to HUT position [21, 22]. Although this analysis is simple, it ignores all of the blood pressure signal and only analyzes the discrete change in heart rate between the supine and HUT position ignoring all features within the signal. The analysis performed in this study was motivated by visual inspection of data, revealing that compared to control subjects, POTS patients display a higher amplitude of 0.1 Hz oscillations.

To our knowledge, only a few previous studies have analyzed the oscillatory behavior of data from POTS patients. One study by Stewart *et al*. [23] describes oscillations in POTS patients using measurements of heart rate, systolic blood pressure, and transcranial doppler measurements of cerebral blood flow velocity. Results from this study using autospectra techniques concluded that cerebral blood flow velocity in POTS patients, all experiencing orthostatic intolerance, oscillated with a larger amplitude as compared to control subjects. Another study by Medow *et al*. [24] investigated the oscillatory dynamics of neurocognition in POTS patients using similar methods as Stewart *et al*. [23]. These studies were able to quantify the amplitude of the 0.1 Hz frequency band but were unable to examine the 0.1 Hz frequency signal with respect to time.

The baroreflex changes the power and instantaneous frequency of both HR and SBP with respect to time in response to physiological changes. Therefore, to analyze the data, it is essential to use methods that can analyze non-stationary and noisy signals, e.g., [25-27]. One popular method for analysis of non-stationary signals is EMD, which has successfully been used to analyze similar data during exercise and HUT [26, 28]. These studies applied EMD to quantify how a change in physiological state (HUT, exercise, or the Valsalva maneuver) affects oscillations in RR intervals and arterial blood pressure. In the present study, we use a similar methodology to quantify the effects of a HUT test in control subjects and POTS patients. By using Uniform Phase Empirical Mode Decomposition (UPEMD), which essentially filters the non-stationary data extracting the 0.1 Hz component of the signal, we can analyze how this portion of the signal changes in time and use stationary methods to analyze the power of the oscillations. Obtaining a signal in the time-domain is advantageous as it can be used to characterize the phase response of the signals both at rest and during HUT.

We use random forest machine learning to calculate the importance of each metric to the correct classification of patients but do not present a classification model for future data due to the limited number of subjects in this study [29]. We compute predictor association and use $k$-means clustering to categorize data based on the developed metrics and traditional diagnostic criteria. We then compare the cluster groupings with the diagnosis of by physicians.

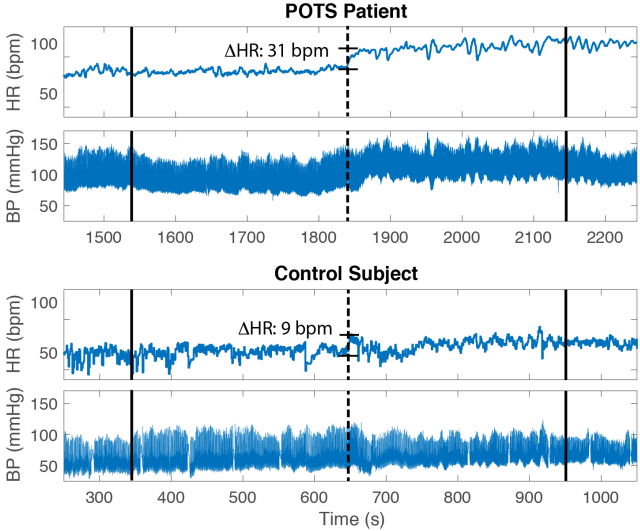

Fig. 1: Example heart rate (HR, bpm) and blood pressure (BP, mmHg) data from a POTS patient (top) and a control subject (bottom). The solid vertical lines indicate the start and end of the data segment analyzed. The dashed vertical line denotes the onset of HUT. For the POTS patient, the resting heart rate is higher (75 bpm) compared to the control subject (50 bpm) and it is increased by 31 bpm. For the control subject heart rate (HR) is increased by only 9 bpm.

## II. METHODS

To characterize oscillations in POTS patients and control subjects, we analyze non-stationary electrocardiogram (ECG) signals and blood pressure data from HUT studies. Using these data, we extract heart rate and systolic blood pressure. This study uses systolic blood pressure as this part of the blood pressure signal is associated with dysfunction of autonomic blood pressure control to a larger degree than the diastolic signal [30]. To determine the frequency content of the signals, we use UPEMD to extract stationary signals, known as Intrinsic Mode Functions (IMFs). To study baroreflex regulation, we target the 0.1 Hz frequency range and examine the frequency spectra of the IMFs using Fast Fourier Transformation (FFT). We then fit a Gaussian curve to the transformation of the 0.1 Hz IMF to compare the power of the signal across groups. We characterize the 0.1 Hz IMF phase response by calculating the average instantaneous phase difference between the 0.1 Hz heart rate and systolic blood pressure IMFs. We then compute the spontaneous baroreflex sensitivity (SBR) for every patient to compare against our phase difference metric. Finally, we use machine learning and clustering to determine what metrics best characterize the two groups (POTS patients vs. control subjects).

### A. Experimental Protocol

Data summarized in Table I are extracted from clinical examinations at Frederiksberg and Bispebjerg Hospitals, Denmark. All data are collected with approval from the Frederiksberg and Bispebjerg Hospitals ethics committee, and all subjects gave written consent to participate in research studies. Data analyzed include ECG and blood pressure measurements from 28 women with a positive POTS diagnosis and 28 female control subjects. This patient group was chosen as POTS primarily affects women [8]. Patients with severe arrythmia, who experienced syncope, or were diagnosed with other cardiovascular or neurological diseases were excluded from this study.

Patients were given a POTS diagnosis if they experienced orthostatic intolerance episodes and exhibited a heart rate increase of more than 30 bpm (40 bpm if aged 12-19 years), or if they maintained a heart rate at or above 120 bpm in upright position [31].

For all patients, the ECG readings were obtained from a precordial ECG-lead, while blood pressure was measured using photoplethysmography on digital arteries in the index finger on the non-dominant hand (Finapres Medical Systems BV, Amsterdam, The Netherlands). The Finometer signal was calibrated against sphygmomanometer measurements. Both signals were sampled at a frequency of 1000 Hz. Deidentified data were stored in LabChart (LabChart, AD Instruments Inc., Colorado Springs, CO, USA).

Patients begin the procedure resting in the supine position for at least three minutes before being tilted head-up to 60º at a speed of 15º per second measured by way of an electronic marker. Subjects remained tilted head-up for at least three minutes. For this study, we extract 6-10 minutes of data, including, up to five minutes before the HUT and 5 minutes during the HUT. This produces up to 600 seconds of data for each patient, as illustrated in Fig. 1 depicting the raw heart rate (bpm) and blood pressure (mmHg) signals for a POTS patient and a control subject, respectively. For all datasets, the HUT maneuver was one of several tests performed to assess the autonomic control system. For all datasets we aimed at including five minutes of data before and during the tilt, as longer data-segments allow for more reliable signal processing results. The study performed here is a retrospective analysis, therefore we were not able to extract five minutes before or during the HUT for all subjects. Shorter supine segments were caused by a delay in initiation of the recording device – meaning the patient was resting but data were not recorded until later in their resting period. Shorter HUT segments were because some other test was performed (e.g. sublingual nitroglycerine, carotid massage). For these patients we included as much data as possible. The supine segment was reduced for 7 patients (for 4 patients we extracted 4-5 minutes segments and for 3 patients we extracted 3-4 minutes segments). The HUT segment was reduced for 5 patients (for 4 patients we extracted 4-5 minutes segments and for 1 patient we extracted 3-4 minutes segments). To ensure that the analysis meet criteria suggested by [7], for all datasets we analyzed a least 3 minutes of data before and during the HUT.

Table I: Standard Patient Characteristics.

| Subject Group | Age (years) | Height (cm) | Weight (kg) | $\Delta$ HR (bpm) |
|---|---|---|---|---|
| POTS | 25.6 ± 10.0 | 172 ± 6 | 66 ± 14 | 31.2 ± 11.7 |
| Control | 40.4 ± 17.0 | 162 ± 6 | 67 ± 13 | 7.7 ± 6.3 |

Standard patient characteristics presented as mean ± standard deviation for both POTS and control subjects. All subjects were female.

The patient data were separated into two parts representing rest and HUT. The rest period is defined as the up to 300 seconds before the marker noting the onset of HUT. The HUT segment begins at the marker, denoting the HUT onset and ends up to 300 seconds after the procedure starts. Heart rate and systolic blood pressure are then extracted from the ECG and continuous blood pressure time series data.

*1) Heart Rate*
Heart rate (shown in fig. 2a) is calculated from the ECG signal as the inverse of the RR interval for each cardiac cycle. Since the time-series signal is non-stationary, we filtered the ECG signal using the `medfilt1` median filter algorithm in MATLAB twice with a 200 and a 600 ms window storing only the stationary components of the QRS complex and the P-waves [32]. Additional drift in the signal was identified and removed using a Savitsky-Golay filter with 150 milliseconds (ms) and an order 5 polynomial. To identify the peaks in the filtered signal, we used MATLAB's peak detection algorithm `find-peaks` with the minimum distance between peaks set to 200 ms. To ensure the identification of the R peaks, the mean of these peaks is used as a minimum peak height for the `findpeaks` algorithm. Next, we compute the distance between the R peaks and use these to calculate the RR (ms) intervals and the heart rate $HR_i = 60/RR_i$ (bpm), where $RR_i$ is the length of the $i^{th}$ RR interval. This calculation gives heart rate at $n-1$ points, where $n$ is the number of time points. RR intervals and heart rate are depicted in Fig. 2. The smooth heart rate signal (shown in Fig. 2b) is obtained by interpolating over these points using a piecewise cubic Hermite interpolating polynomial (PCHIP) and then subsampling the signal to 250 Hz.

*2) Systolic blood pressure*
Blood pressure (shown in Fig. 2c) is measured continuously using the FinaPres. From this signal, we use the function `findpeaks` in MATLAB with a minimum peak prominence of 25 mmHg and a minimum peak distance of 0.25 seconds to extract systolic blood pressure within each cardiac cycle. Similar to heart rate, we obtain a continuous signal (shown in Fig. 3) by interpolating the discrete signal using PCHIP and then subsampling to 250 Hz.

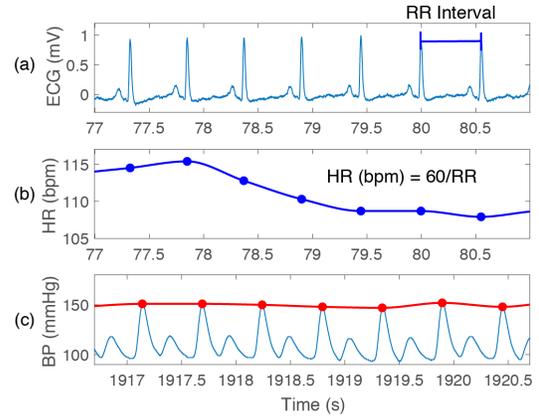

Fig. 2: Snapshot of (a) electrocardiogram (ECG, mV), (b) heart rate (HR, bpm), and (c) blood pressure (BP, mmHg) signals at rest, depicted over a 3-second interval. Beat-to-beat heart rate (HR, bpm) (a) and systolic blood pressure values (red) (c) values are predicted using a peak detection algorithm (red and blue circles). Continuous signals are obtained using PCHIP interpolation.

### B. Uniform Phase Empirical Mode Decomposition

We use UPEMD (an extension of EMD) to analyze the 0.1 Hz frequency response in non-stationary heart rate and systolic blood pressure time series. We chose this method over other methods such as Ensemble EMD (EEMD) [33] since UPEMD has the unique advantage of explicitly targeting a frequency band to be examined [20]. This feature is essential for the analysis of the heart rate and blood pressure data, which have significant frequency signatures in bands close to the 0.1 Hz band, in particular from respiration (~0.25 Hz). This allows us to examine the contribution of the baroreflex (0.1 Hz) with minimal input from other frequencies.

*1) Empirical Mode Decomposition*
EMD [34], decomposes a non-stationary oscillatory signal into a number of stationary IMFs and a residual. The EMD analysis (Algorithm 1) relies on an iterative method, which sifts out the non-stationary portion of the signal, resulting in a stationary oscillatory signal, the *intrinsic mode function* (IMF.). As outlined in Algorithm 1, we find the maxima and minima in the signal, and use these to construct an upper and lower envelope, which we subtract from the data. We repeat this process until it is not possible to obtain more IMFs.

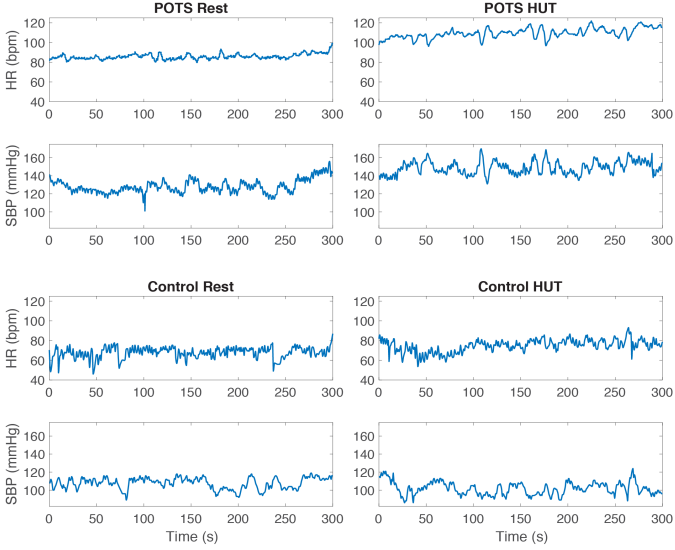

Fig. 3: HR and SBP signals for a sample patient during rest and HUT for a POTS patients (top) and a control subject (bottom). For each patient, we analyze 5 minutes of data at rest and after the onset of HUT.

---

**Algorithm 1 EMD** (Adapted from [34, 35])

**Input:** Signal, $x(t)$
**Output:** IMFs

1    $r_0(t) \leftarrow x(t)$
2    $i \leftarrow 0$
3    **while** $r_i(t)$ does not meet stopping criterion
4      $h_{i,0}(t) \leftarrow r_i(t)$
5      $j \leftarrow 0$
6      **while** $h_{i,j}$ does not meet the numerical IMF criterion
7        Compute upper and lower envelope of $h_{i,j}$, $u_{i,j}$ and $l_{i,j}$ (using cubic splines)
8        $m_{i,j} = (u_{i,j} + l_{i,j})/2$
9        $h_{i,j+1} = h_{i,j} - m_{i,j}$
10       $j \leftarrow j + 1$
11      **end**
12      $c_i \leftarrow h_{i,j}$    ($c_i$ is an IMF)
13      $r_{i+1} = r_i - c_i$
14      $i \leftarrow i + 1$
15    **end**
16    **Return** Matrix with column $n$ equal to IMF $c_n$

---

The IMFs are stationary decompositions of the signal. They have an equal number of maxima and minima, and the number of peaks and troughs differ by at most one. The upper and lower envelopes of the filtered signal defined by the maxima and minima must average to zero at all points [34]. In this study, IMFs are computed using the emd function in MATLAB's signal processing toolbox. This algorithm, described in detail by Huang et al. [34] uses a Cauchy type criterion, that represents the standard deviation (SD) of two consecutive siftings, defined as

$$\text{SD} = \sum_{t=0}^{T} \left[ \frac{\left| h_{i,(k-1)}(t) - h_{i,k}(t) \right|^2}{h_{1(k-1)}^2(t)} \right] < 0.2. \tag{1}$$

As suggested in [34], we impose SD < 0.2. Therefore, we restrict SD to 0.2 for two consecutive siftings, that is, if SD < 0.2. Then $h_{i,k}$ is labeled as the next IMF, $c_i$.

We repeat the sifting until either the residual signal $r_i(t) = x(t) - \sum_{i=0}^{i-1} c_i$ is monotonic and therefore cannot produce more IMFs, or if the energy ratio

$$\text{ER} = 10 \log_{10} \left( \frac{\|x(t)\|^2}{\|r_i(t)\|^2} \right) > \gamma,$$

where $x(t)$ denotes the original signal and $\gamma = 20$ denotes the default energy ratio (ER) threshold. Intuitively, the ER compares the energy of the signal at the beginning of the sifting with the average envelope energy.

Finally, by combining the IMFs and the final residual, it is possible to reconstruct the original signal as

$$x(t) = \sum_{i=1}^{p} c_i + r_{p+1}. \tag{2}$$

*2) Uniform Phase Empirical Mode Decomposition*

A limitation of the EMD method is a phenomenon known as mode mixing referring to IMFs that overlap in the frequency domain or encode vastly different portions of the frequency spectra [36]. To minimize mode mixing, we use the Uniform Phase EMD (UPEMD) [20].

UPEMD (Algorithm 2) averages the IMFs computed with the EMD on a series of perturbed signals. These perturbations are sinusoidal functions that are uniformly distributed on the interval $[0, 2\pi)$. Perturbing the original signal in this way reduces the effects of noise and allows for a more accurate representation of a target frequency free from mode mixing. As suggested in the literature [20], we assume that the number of perturbations $n_p = 16$, the number of IMFs, $n_{imf} = \log_2 n \approx 16$, where $n$ is the number of observation points for a data set, and the target frequency $f_w = 0.1$ Hz.

*3) Analyzing the Power of Intrinsic Mode Functions*

The output of the targeted UPEMD is a collection of IMFs that represent unique frequencies of the original signal in the time domain. Note that, by definition, the IMFs are stationary, and therefore we can compute the one-sided power spectrum of the 0.1 Hz IMF; we use MATLAB's FFT algorithm.

The FFTs of the IMFs with mean frequencies 0.05-0.5 Hz are shown in Fig. 5 for the two characteristic data sets. To determine the power of the 0.1 Hz frequency response across the population, we fit a Gaussian distribution function $f(\omega)$, to the data of the form

$$f(\omega) = a e^{-\left(\frac{\omega - b}{c}\right)^2}, \tag{3}$$

where $a$ is the maximum amplitude of the Gaussian function, $b$ is the value at which the function achieves its' maxima, and $c$ contributes to curve width. Fig. 6 shows the FFT and Gaussian fits for the two characteristic subjects. To determine differences between position and disease (POTS), we compare values of $a$, the amplitude of the Gaussian fitted to the 0.1 Hz spectra of the IMF.

---
**Algorithm 2** UPEMD (Adapted from [20])
**Input:** Signal, $x(t)$
**Output:** $n_{imf}$ IMFs

1      $r_0(t) \leftarrow x(t)$
2      **for** n = 1 to $n_{imf}$
3          $\epsilon_n = std(r_{n-1}(t))$
4          **for** k = 1 to $n_p$
5              $y_k(t) = r_{m-1} + \epsilon \cos\left(2\pi\left(f_w t + \frac{k-1}{n_p}\right)\right)$
6              $c_{n,k}(t) = first$ column of EMD $(y_k(t))$
7          **end**
8          $c_n(t) = \frac{1}{n_p} \sum_{k=1}^{n_p} c_{n,k}(t)$
9          $r_n(t) = r_{n-1}(t) - c_n(t)$
10     **end**
11     **Return** (matrix with column $n$ equal to $c_n$)

---

*4) Quantification of Phase Dependence*
The afferent baroreceptor nerves sense changes in blood pressure. The signal is transmitted to the brain via negative feedback, mediating changes in HR, vascular resistance, compliance, and cardiac contractility. Hence, the analysis of the interaction between the two signals gives additional insight into the baroreflex function. To quantify the responsiveness of the baroreflex, we examine the interaction of the phases of the 0.1 Hz IMF for the heart rate and blood pressure signals at every time point. The baroreflex responds to an increase in blood pressure by decreasing heart rate, and a decrease in blood pressure by increasing heart rate [37]. This implies that the baroreflex is a negative feedback loop, and therefore, a phase difference of $\pi$ implies that the reflex is instantaneous. To quantify the responsiveness of the baroreflex, we calculate the relative difference between $\pi$ and the instantaneous phase difference between the two 0.1 Hz signals. To do so, we utilize that the properties of IMFs allow the application of the Hilbert Transform to calculate the instantaneous phase [34].

Let $Z(t)$ denote the IMF, $HT[Z(t)]$ the Hilbert Transform of $Z(t)$, the instantaneous phase $\hat{\phi}(t)$ is then given by

$$\hat{\phi}(t) = \tan^{-1}\left(\frac{HT[Z(t)]}{Z(t)}\right) \quad (4)$$

[38]. We compute a continuous version of the instantaneous phase by using the `unwrap` command in MATLAB, denoted here by $U(X(t))$. This gives a continuous instantaneous phase, $\phi(t)$, defined as

$$\phi(t) = U\left(\frac{HT[Z(t)]}{Z(t)}\right). \quad (5)$$

For each data set, we denote the instantaneous phase of the 0.1 Hz heart rate IMF by $\phi_{HR}(t)$, and the instantaneous phase of the 0.1 Hz systolic blood pressure IMF as $\phi_{SBP}(t)$. Defining $T$ as the length of the signal in seconds, we quantify the interaction of the two signals by

$$M_h = \frac{1}{T}\int_0^T \left|mod_{2\pi}(\phi_{HR}(t) - \phi_{SBP}(t)) - \pi\right| dt. \quad (6)$$

This equation quantifies the average distance of the instantaneous phase difference from $\pi$, the instantaneous baroreflex. A value of $M_h = 0$ implies that as systolic blood pressure increases/decreases, heart rate compensates by decreasing/increasingly instantaneously. Hence, a smaller value of $M_h$ ($0 \leq M_h < \pi$) represents a more responsive, baroreflex. Our assumption that the period of these oscillations is approximately 10 seconds implying that $M_h = \pi$ corresponds to a response time of 5 seconds [39]. The bounds of $M_h$ therefore agree with the current understanding of the baroreflex [40]. We calculate the relative difference between $\pi$ and the instantaneous phase difference between the 0.1 Hz frequency component of the signals.

*5) Spontaneous Baroreflex Sensitivity (SBR)*
The spontaneous baroreflex sensitivity (SBR) quantifies the change in HR due to the change in blood pressure. To calculate SBR, we determine the mean slope of a regression line through three or more consecutive systolic blood pressure peaks that are either increasing or decreasing when plotted against the RR interval of the beat following the systolic blood pressure peak [41].

*6) Statistical Analysis*
To compare the power of the 0.1 Hz frequency in our data, and the phase responses, we use the one-way Analysis of the Variance function `ANOVA1` in MATLAB.

*7) Random Forest and Clustering Analysis*
Given the eight metrics identified in this study, we seek to cluster the data to quantify the importance of each metric to correct grouping and how patients are grouped based on multiple diagnostic metrics. We first find the most important metrics by classifying patients using random forest machine learning. We use the MATLAB function `fitcensemble` to create an ensemble of 100 trees and compute the k-fold loss ($k=10$) for different maximum branching numbers to prevent overfitting. We then compute the predictor importance using the function `oobPermutedPredictorImportance`. We then employ $k$-means clustering to classify the data, including new metrics as well as the change in heart rate from supine to HUT and average HR during HUT [42].

## III. RESULTS

For each patient, our analysis produces an IMF that represents the 0.1 Hz frequency of the signal with respect to time for heart rate and systolic blood pressure at both rest and HUT, totaling 4 IMFs per patient. The 0.1 Hz IMFs for one POTS patient and one control subject are shown in Fig. 4. We compare the power and phase difference of the signals across groups and use random forests and clustering to determine the importance of metrics.

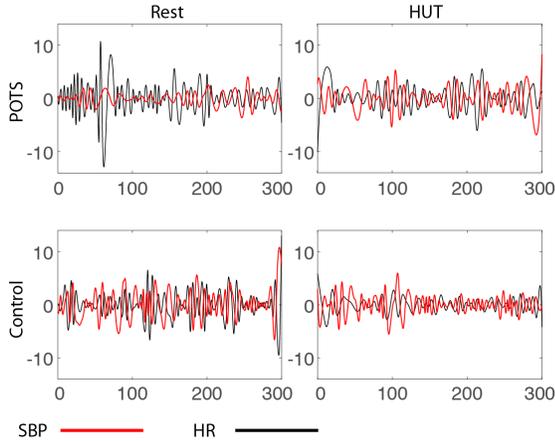

Fig. 4: Systolic blood pressure (SBP, mmHg - red) and heart rate (HR, bpm - black) IMF's containing 0.1 Hz oscillations at rest (left) and during HUT (right) for a representative POTS patient (top) and control subject (bottom).

### 1) Signal Power

Each group contains predictions from 28 subjects. The results of the FFT of the IMFs are shown in Fig. 5 for a representative control subject and POTS patient. This figure shows that the frequencies cluster at 0.1 Hz characterizing the power of the baroreflex response [17]; 0.25 Hz characterizing respiration and the response of the RSA reflex [13], and a broad distribution at higher frequencies (~0.3–0.5 Hz); the last frequency distribution (yellow) is wide and nearly uniform, and therefore most likely shows noise in the original signal. The results of applying a Gaussian fit of the FFT of the 0.1 Hz IMF are shown in Fig. 6 for 2 subjects. For each subject, the amplitude $a$ (reported in Table II) of the 0.1 Hz frequency response is computed as the max of the Gaussian distribution.

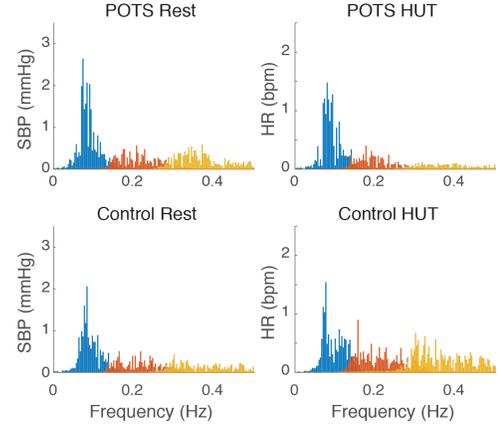

Fig. 5: Amplitude of oscillations of the frequency bands detected by UPEMD. The amplitude was computed from FFT of the IMFs. The blue spectra show the FFT of the 0.1 Hz IMF; orange spectra show the FFT of the respiratory frequency (~0.2 Hz) IMF; the broad yellow spectra show the FFT of the high-frequency IMF.

Table II: Signal characteristics.

|   | Ctrl Rest | Ctrl HUT | POTS Rest | POTS HUT |
|---|---|---|---|---|
| $a_{HR}$ | $0.55 \pm 0.31$* | $0.52 \pm 0.31$* | $0.82 \pm 0.27$* | $1.03 \pm 0.43$* |
| $a_{SBP}$ | $0.71 \pm 0.22$* | $0.86 \pm 0.36$* | $0.68 \pm 0.20$* | $1.22 \pm 0.49$* |
| $M_h$ | $1.29 \pm 0.24$* | $1.31 \pm 0.30$* | $0.95 \pm 0.25$* | $0.90 \pm 0.31$* |
| $\overline{HR}$ | $72 \pm 12$ | $78 \pm 12$* | $74 \pm 13$ | $104 \pm 16$* |
| $\overline{SBP}$ | $117 \pm 21$ | $124 \pm 22$ | $110 \pm 16$ | $112 \pm 16$ |
| $\Delta HR$ |  | $7.6 \pm 5.8$ * |  | $32.0 \pm 11.7$* |
| SBR | $6.36 \pm 4.7$ | $3.81 \pm 2.35$ | $8.09 \pm 4.61$ | $3.17 \pm 1.92$ |

Numbers are reported as the mean $\pm$ standard deviation. $\Delta HR$ report the change in heart rate from rest to HUT. A * marking denotes that the marker is used in the random forest and clustering analysis.

Table III: One-way ANOVA comparing the amplitude of the 0.1 Hz IMF oscillations.

| Comparison |  |  | p - value |
|---|---|---|---|
| Rest vs HUT | Control | HR | 0.69 |
|  |  | SBP | 0.06 |
|  | POTS | HR | 0.03 |
|  |  | SBP | < 0.005 |
| Control vs POTS | Rest | HR | < 0.005 |
|  |  | SBP | 0.57 |
|  | HUT | HR | < 0.005 |
|  |  | SBP | < 0.005 |

ANOVA compressions of amplitude of 0.1 Hz component of various signals. We use 0.005 as our threshold for statistical significance.

Results show that in POTS patients, the amplitude of systolic blood pressure 0.1 Hz oscillations is significantly larger during HUT than at rest, but we fail to reject the null hypothesis for the same comparison in the control subjects for both HR and SBP, and for HR in POTS patients. The amplitude of the heart rate oscillations is larger in POTS patients compared to control subjects both at rest and during HUT, and the amplitude of the systolic blood pressure oscillations is only larger between control subjects and POTS patients during HUT. These results indicate that at rest, the sympathetic branch can maintain blood pressure at homeostasis, while the parasympathetic branch is impaired both at rest and during HUT.

Figure 7 shows box plots comparing the 0.1 Hz amplitude for each group. ANOVA tests, summarized in Table III, compare predictions of $a$ (maxima of the Gaussian fit of the 0.1 Hz Fourier spectra) between the four groups: Rest (control subjects and POTS patients) and HUT (control subjects and POTS patients). Overall, the results presented here indicate that the POTS patients exhibit an abnormally sensitive baroreflex when compared to the control subjects.

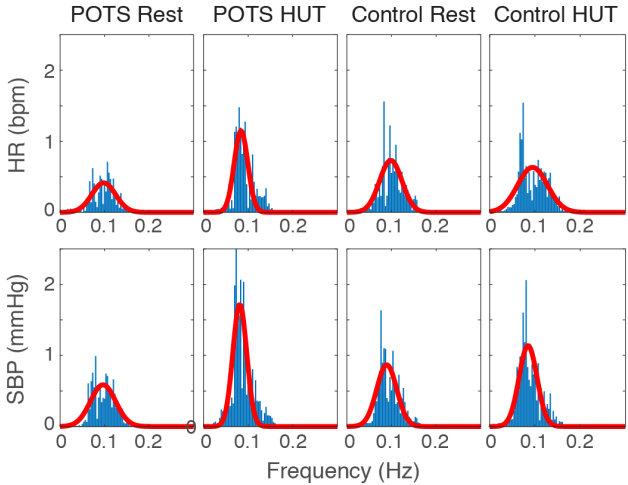

Fig. 6: Gaussian fit to 0.1 Hz heart rate (HR, top) and blood pressure (BP, bottom) spectra for a representative POTS patient (left) and control subject (right). The amplitude of the power is summarized in Table IV averaging the response for all subjects in each group.

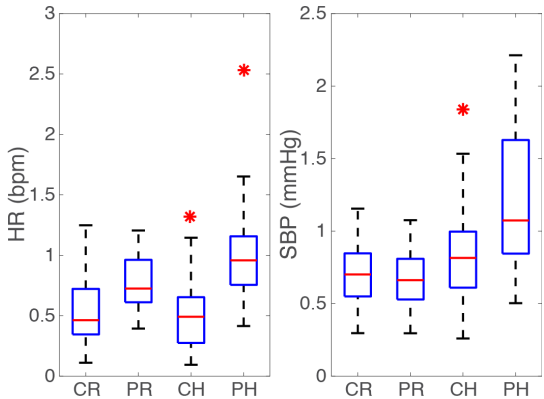

Fig. 7: Box and whisker plots, comparing the amplitude of the 0.1 Hz IMFs for each patient group: controls at rest (CR), POTS at rest (PR), controls during HUT (CH), and POTS during HUT (PH). A red * denotes an outlier.

### 2) Phase Response

To compare the instantaneous phase difference $M_h$ across groups, we performed an ANOVA analysis, including predictions from 28 subjects per group. Calculated values of $M_h$ are reported in Table II and illustrated in Fig. 8. The ANOVA analysis (summarized in Table III) comparing predictions between groups show that $M_h$ is significantly smaller in POTS patients compared to controls, both at rest and during HUT, but it does not change significantly between rest and HUT within control subjects or POTS patients. The decreased $M_h$ value in POTS patients implies that they have a faster baroreflex response than the control subjects regardless of their orthostatic position.

We conduct the same comparisons for the action (SBR) calculated with a 1 heartbeat delay, as is traditionally done. Results of SBR are reported in Table II, and ANOVA of SBR are presented in Table III. The only significant difference for SBR is in POTS patients between rest and HUT. We also observe that the coefficient of variation (Standard deviation divided by mean) of SBR is greater than 50% for all groups, whereas the coefficient of variation of $M_h$ is below 25%.

Table III: One-way ANOVA Comparisons for $M_h$.

| Comparison | $M_h$ p - value | SBR p-value |
| --- | --- | --- |
| Control rest   vs.  Control HUT | 0.82 | 0.01 |
| POTS rest      vs.  POTS HUT | 0.59 | < 0.005 |
| Control rest   vs.  POTS Rest | < 0.005 | 0.17 |
| Control HUT   vs.  POTS HUT | < 0.005 | 0.27 |

$p$-values from a one-way ANOVA comparing $M_h$. We use 0.005 as our threshold for statistical significance.

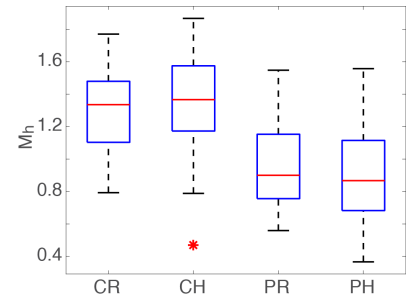

Fig. 8: Box and whisker plots comparing the instantaneous phase difference ($M_h$) for each patient group: controls at rest (CR), controls during HUT (CH), POTS at rest (PR), and POTSs at HUT (PH). A red * denotes an outlier.

### 3) Clustering Analysis

We used random forest machine learning to determine what factors provided better predictors for POTS. We compared eight predictors given in Table II, including the amplitude of the 0.1 Hz oscillations of heart rate during rest/HUT, systolic blood pressure during rest/HUT, the change in heart rate during HUT, and the average heart rate during HUT. Note, that average heart rate during HUT is calculated over the entire extracted HUT segment. The value of k-fold loss for different maximum branching numbers varies from 0.04 – 0.08, with a value of 0.07 with a maximum branching number of 8 (number of predictors used). Results in Fig. 9a shows that the four most important metrics to detect POTS include: (1)

the change in heart rate ($\Delta H$) between rest and HUT, (2) the average heart rate during HUT (Hm) (3) the phase difference $M_h$ at rest (MR), and (4) the amplitude of heart rate oscillations during HUT (HaH). Subsequently, we used clustering with $k$-means to cluster with all eight metrics. Figure 9b shows the silhouette plot of the predicted clusters. This plot shows how similar a member of a cluster is to other members of the same cluster. The silhouettes have an average length of 0.47. The clustering labeled three patients previously diagnosed POTS patients as control, and two controls as POTS patients.

The machine learning analysis assumes that the medical diagnosis is accurate, which may be true. All patients underwent a series of tests, including a Valsalva maneuver, a head-up tilt test, a deep breathing test, and an active standing test. In principle, all tests should result in a POTS type response, but in practice, some tests may fail to do so.

In summary, our results show that the amplitude of the 0.1 Hz frequency component of the heart rate and systolic blood pressure is larger in POTS patients during HUT and that the phase response between heart rate and systolic blood pressure is shorter in POTS patients. Machine learning and clustering analysis show that the phase difference at rest is an effective metric that can be calculated at without subjecting the patients to HUT.

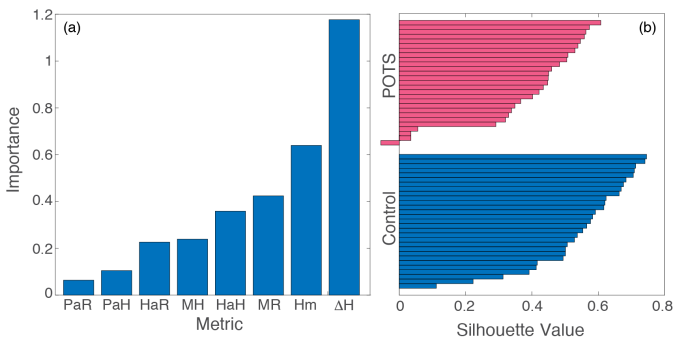

Fig. 9: (a) Importance of computed metrics, including the change in heart rate between supine and HUT ($\Delta H$), the mean heart rate during HUT (Hm), the phase difference between blood pressure and heart rate at rest (MR), the amplitude of the 0.1 Hz heart rate oscillations during HUT (HaH), the phase difference between heart rate and blood pressure during during HUT (MH), the amplitude of the 0.1 Hz heart rate oscillations at rest (HaR), the amplitude of the 0.1 Hz blood pressure oscillations during HUT (PaH), and the amplitude of the 0.1 Hz blood pressure oscillations at rest (PaR). (b) clustering of POTS patients (top) and control subjects (bottom). Three POTS patients were classified as controls and two control subjects as POTS patients.

## IV. DISCUSSION

Our study has shown that postural orthostatic tachycardia (POTS) patients have a more pronounced 0.1 Hz frequency response compared to controls both during rest and head-up-tilt (HUT). This frequency response is associated with the baroreflex [43]. Therefore, our results indicate, as hypothesized, that POTS patients have an oversensitive baroreflex causing significant and rapid changes in both systolic blood pressure and heart rate.

POTS is currently diagnosed by the presence of tachycardia without syncope upon postural change from supine to standing or with HUT. Positive diagnosis requires a change in heart rate of more than 30 bpm in adults (40 bpm in children/young adults aged 12 to 19 years old). The patient data analyzed here were categorized as POTS if they experienced orthostatic intolerance and a significant increase in heart rate or a sustained high heart rate (above 120 bpm) during HUT or an active standing test. The analysis performed here focuses on identifying markers that correlate with the baroreflex response (~0.1 Hz), enabling us to generate physiological hypotheses explaining the observed oscillations in heart rate and blood pressure [22], i.e., the $\Delta$HR response was not included in the analysis. The overactive system could correlate to findings by us (not published) and others [44], noting that most POTS subjects express agonistic antibodies that bind to cardiac pacemaker cells and smooth muscle cells within the arterial wall. While the presence of specific autoantibodies does not confirm disease causality, and results are difficult to translate to system-level blood pressure and heart rate observations, the correlation between these observations suggests that the baroreflex system may be compromised in this patient subgroup.

By using UPEMD [20], we can represent the nonstationary signals as a series of stationary components that can be analyzed using stationary methods in both the time and frequency domain. Our results show that POTS patients exhibit larger 0.1 Hz oscillations in both heart rate and systolic blood pressure both at rest and during HUT. This result agrees with previous studies, e.g., the study by Stewart et al. [23], which quantified the amplitude of cerebral blood flow and blood pressure oscillations. Results from that study suggested that these oscillations may be responsible for decreased neurocognition "brain fogginess" in POTS patients.

This study analyzed data from a HUT, but other tests could be used including the Valsalva maneuver and active standing. The HUT test is a common diagnostic procedure for patients with syncope [45] but is here used for analysis of patients with POTS, where pathological changes occur immediately upon assumption of the upright posture, in contrast to syncope that develops over longer period of time. The use of 6-10 min of data for the analysis agrees with [7].

The advantage of UPEMD is that we are able to treat the signals as non-stationary, which is motivated by our understanding of the baroreflex and cardiovascular system. Previous studies of these signals have used stationary techniques as approximations; however, we elect to use UPEMD in this study to agree with theory. Using UPEMD, we are also able to study the frequency response in time (as shown in Fig. 8). This allows us to quantify the phase relationship (via $M_h$) between the signals, a novel result that to our knowledge has not been reported previously.

Our analysis compared eight metrics: the amplitude of 0.1 Hz oscillations at rest and HUT for both HR and BP; the phase difference between heart rate and blood pressure at rest and during HUT; the change in heart rate during HUT, and the magnitude of heart rate during HUT. Results of the random forest analysis revealed that in addition to the change in heart rate, the phase difference (our new marker) between the two signals provides the most significant markers. By using a maximum branching number of 8 (number of predictors) we obtained a k-fold loss of ~0.07. We view this as acceptable for our purpose of ranking importance of metrics. The instantaneous phase difference is of importance as it provides a new way to quantify baroreflex sensitivity other than the spontaneous baroreflex (SBR) method.

The new marker, $M_h$, quantifies the response time of the baroreflex, whereas SBR measures the magnitude of the baroreflex response. We see from our analysis that SBR changes in POTS patients when transitioned from rest to HUT. This implies that SBR can only detect abnormal baroreflex activity in POTS patients with a HUT, whereas $M_h$ can detect an abnormal baroreflex during rest and HUT. Furthermore, since $M_h$ does not change from rest to HUT, and SBR does for POTS patients, we argue that $M_h$ is better at detecting abnormal responses.

A potential problem with SBR is the assumption that the response of a change of pressure can be fully quantified by the RR interval of the next heartbeat. In reality, this response time is likely not equal to the next RR interval time. Our metric, $M_h$, represents this response time, and is calculated in a continuous fashion. Hence, future work could calculate a continuous version of SBR using a patient specific response time ($M_h$).

The new marker, $M_h$, is calculated using the Hilbert Transform to find the instantaneous phase of this signal. It should be noted that one could calculate the instantaneous amplitude of the signal via the Hilbert Transform to characterize the amplitude of the ~0.1 Hz oscillations and obtain similar results as using the FFT method presented above.

A diagnosis of POTS is made using a number of criteria: that patients showed signs of orthostatic intolerance (a metric not directly quantifiable by the data analyzed), that they exhibited an increase in heart rate upon standing, in response to a Valsalva maneuver (data not analyzed), a HUT (analyzed here), or that they had a sustained heart rate at or above 120 bpm. Clustering analysis characterizing POTS by a ΔHR > 30 or a sustained heart rate of more than 120 bpm resulted in 6 POTS patients classified as control subjects. Neither of these patients had a ΔHR > 30 bpm.

Nevertheless, an inspection of data from Valsalva maneuvers and active standing tests showed that tests were associated with a heart rate increase ΔHR > 30 bpm. In comparison, classification, including the 0.1 Hz frequency response metric, identified only three POTS patients as controls (also labeled control if only ΔHR was considered). These patients did not experience a change in heart rate at or above 30 bpm, but they all exhibited a fairly high heart rate response to active standing ($\Delta HR \approx 30$). It should be noted that one of the three miscategorized patients had a very high resting heart rate, and almost no oscillations. We hypothesize that this patient may have POTS combined with inappropriate sinus tachycardia.

The clustering analysis performed in this study placed all POTS patients in one group, even though the phenotype likely are caused by different mechanisms [6]. The lack of differentiation within the POTS group could be because all POTS patients analyzed had the same genotype, that the sample size was too small, or that the HUT tilt is not able to differentiate the sub-groups. The approach used here can easily be expanded by including a larger sample size and by comparing markers identified by the HUT test to other tests, e.g. the Valsalva maneuver or active standing.

In addition, two control patients were categorized as POTS. These patients could have been misdiagnosed. Most of our data from control subjects are from people contacting the autonomic clinic because they experienced orthostatic intolerance but were classified as healthy since their heart rate response did not display abnormal features according to existing protocols. Overall, our results are promising, and they motivate future work. In particular, it would be beneficial to include data from other tests including active standing and the Valsalva maneuver.

This study is limited as we only analyze data from 56 patients (28 POTS patients and 28 control subjects). Due to this limitation, we were not able to match patients based on demographics. Future studies should include more datasets, potentially including more measurements per patient, including demographics and orthostatic intolerance markers.

Another disadvantage of this limited sample size is the inability to validate the model produced by Random Forest Machine Learning on an independent cohort of patients. For this reason, the Random Forest approach is only used to calculate the importance of the metrics presented in this paper and cannot be used as a diagnostic tool. Future studies, with more data sets, should use machine learning to create, and validate, models that can assist with clinical diagnosis.

In summary, we present evidence that heart rate and blood pressure oscillations are essential to understanding the underlying dynamics of POTS and provide a way to incorporate the detection of oscillations into the diagnosis protocol. We argue that by quantifying both the oscillations and an increase in heart rate, clinicians will be able to provide a more accurate patient diagnosis. We showed that in addition to changes in heart rate, POTS diagnosis should include metrics computing the amplitude of the heart rate and systolic blood pressure 0.1 Hz frequency response and the phase difference between the heart rate and blood pressure signals. These metrics all agree with our hypothesis that the baroreflex is enhanced in POTS patients. The addition of our new metrics comparing the heart and blood pressure response opens an avenue providing more insight into the pathophysiology of POTS. POTS is typically a comorbidity in a number of conditions, including visceral pain, chronic fatigue [22], migraine, joint hypermobility [46], and chronic anxiety [44]. Including the specific comorbidity, and our new POTS markers may allow us to differentiate between the POTS patients, essential to generate better treatment protocols.

### III. CONCLUSION

This study demonstrates that the amplitudes (power) of heart rate and blood pressure oscillations are increased and that the instantaneous phase difference between heart rate and blood pressure is smaller in POTS patients compared to controls. The amplitude of the 0.1 Hz response of HR during HUT and the instantaneous phase difference both at rest and HUT are the most significant markers for POTS. This result indicates that POTS patients have a hypersensitive baroreflex even at rest, indicating that it may be possible to diagnose POTS without invoking the HUT test. We speculate that these oscillations may be responsible for symptoms of the disease, in particular, fatigue as the body uses excessive energy to keep blood pressure at homeostasis. Based on our findings, we suggest that POTS diagnosis protocols should characterize oscillations at 0.1 Hz, providing a more detailed insight into the disease pathophysiology, e.g., by differentiating between tachycardia caused by a reduced central blood volume as opposed to increased baroreceptor sensitivity.

### ACKNOWLEDGEMENTS

This work was supported in part by the National Science Foundation under awards: NSF/DMS(RTG) #1246991 and NSF/DMS #1557761.

We would like to thank Dr. Kun Hu, Division of Sleep Medicine, Harvard Medical School who supplied the preliminary UPEMD code and Dr. Pierre Gremaud, Department of Mathematics, North Carolina State University who assisted with the machine learning and clustering portions of this study.